**Compositional dependence of the direct and indirect band gaps in Ge$_{1-y}$Sn$_y$ alloys from room temperature photoluminescence: Implications for the indirect to direct gap crossover in intrinsic and *n*-type materials**

L. Jiang[1], J.D. Gallagher[1], C.L. Senaratne[2], T. Aoki[3], J. Mathews[4], J. Kouvetakis[2], and J. Menéndez[1]

[1]Department of Physics, Arizona State University, Tempe, AZ 85287-1504
[2]Department of Chemistry and Biochemistry, Arizona State University, Tempe, AZ 85287-1604
[3]LeRoy Eyring Center for Solid State Science, Arizona State University, Tempe, AZ 85287-1704
[4]Department of Physics, University of Dayton, Dayton, OH 45469-2314

**ABSTRACT**

The compositional dependence of the lowest direct and indirect band gaps in Ge$_{1-y}$Sn$_y$ alloys has been determined from room-temperature photoluminescence measurements. This technique is particularly attractive for a comparison of the two transitions because distinct features in the spectra can be associated with the direct and indirect gaps. However, detailed modeling of these room temperature spectra is required to extract the band gap values with the high accuracy required to determine the Sn concentration $y_c$ at which the alloy becomes a direct gap semiconductor. For the direct gap, this is accomplished using a microscopic model that allows the determination of direct gap energies with meV accuracy. For the indirect gap, it is shown that current theoretical models are inadequate to describe the emission properties of systems with close indirect and direct transitions. Accordingly, an *ad hoc* procedure is used to extract the indirect gap energies from the data. For $y < 0.1$ the resulting direct gap compositional dependence is given by $\Delta E_0 = -(3.57\pm0.06)y$ (in eV). For the indirect gap, the corresponding expression is $\Delta E_{ind} = -(1.64\pm0.10)y$ (in eV). If a quadratic function of composition is used to express the two transition energies over the entire compositional range $0 \leq y \leq 1$, the quadratic (bowing) coefficients are found to be $b_0 = 2.46\pm0.06$ eV (for $E_0$) and $b_{ind} = 0.99\pm0.11$ eV (for $E_{ind}$). These results imply a crossover concentration $y_c = 0.073^{+0.007}_{-0.006}$, much lower than early theoretical predictions based on the virtual crystal approximation, but in better agreement with predictions based on large atomic supercells.

## I. INTRODUCTION

The growth of α-Sn films on InSb and CdTe substrates and the epitaxial stabilization of their diamond structure beyond the bulk α→β transition temperature led to the speculation that a similar synthetic route might enable the growth of $Ge_{1-y}Sn_y$ alloy films,[1] thereby bypassing the very limited miscibility of the Sn-Ge system. The interest in the $Ge_{1-y}Sn_y$ alloy stems from the observation that an interpolation of the band structures of its end members Ge and α-Sn suggests that the material should be a direct gap semiconductor over a broad compositional range $y > 0.2$. This estimate was confirmed by microscopic calculations within the virtual crystal approximation (VCA).[2,3] While several III-V and II-VI compound systems feature direct band gaps over the same energy range, the very weak ionic character of the Ge-Sn bond guarantees the near absence of polar phonon scattering, which in III-V and II-VI systems is one of the major factors that limit carrier mobility. Moreover, while the integration of zincblende materials with Si substrates is problematic due to the appearance of anti-phase domains, such domains are expected to be absent in $Ge_{1-y}Sn_y$ alloys due to their average diamond structure.

The experimental confirmation of the above ideas has proven extremely difficult, both from the synthetic and optical characterization standpoints. The first optical studies of the direct-indirect transition in $Ge_{1-y}Sn_y$ alloys were performed in samples grown by low energy ion-assisted Molecular Beam Epitaxy (MBE).[4] The direct and indirect gaps were extracted from optical absorption measurements. To obtain good agreement with experiment, an Urbach tail contribution, representing localized states, had to be added to the model.[4] The results indicated that the compositional dependence of the direct gap deviates strongly from the predicted nearly linear interpolation between Ge and α-Sn. The non-linear dependence was found to be well described by a term of the form $-by(1-y)$, with a bowing parameter $b = 2.8$ eV. For the indirect gap, the deviations from linearity were found to be less pronounced (a finding that was later justified theoretically[5]) implying that the crossover composition is much less than $y_c = 0.2$.

The significant bowing in the compositional dependence of the direct gap has since been confirmed by several groups. However, the reliability of the indirect gap energies determined from absorption measurements is not firmly established. This is because the direct and indirect edges have very different strengths—by about two orders of magnitude— while being in close proximity to each other. This poor contrast is exacerbated by alloy broadening. In fact, ellipsometric studies of the near-band gap optical properties in $Ge_{1-y}Sn_y$ alloys show that the energy dependence and magnitude of the optical absorption is in excellent agreement with calculations that assume a broadened direct edge, without inclusion of indirect or Urbach contributions.[6,7] These studies were performed on $Ge_{1-y}Sn_y$ alloys grown by chemical vapor deposition (CVD) using $Ge_2H_6$ and $SnD_4$ as Ge and Sn precursors, respectively.[8] The CVD method yielded films that not only could be deposited directly on Si substrates, achieving the goal of integrating the $Ge_{1-y}Sn_y$ system with Si technology, but had a high level of crystalline perfection, as demonstrated from direct structural studies, from the fabrication of diodes with good electrical and optical characteristics,[7,9,10] and from the observation of band gap photoluminescence (PL) and electroluminescence.[11,12]

The room-temperature PL signal from $Ge_{1-y}Sn_y$ alloys shows contributions from the direct and indirect band gaps. The simultaneous observation of emission from the two edges is a unique property of Ge-like materials, resulting from the very small (~0.1 eV) separation between them. In bulk Ge, the higher-energy direct gap emission signal is largely suppressed by self-absorption,[13] but it becomes very prominent in films with thicknesses on the order of a few microns or less.[14-18] The observation of distinct direct and indirect gap peaks in PL spectra makes this technique a superior alternative to absorption for the determination of the gap energies. However, while in most low-temperature PL studies of semiconductors the identification of the peak maximum with the associated band gap energy is sufficient for practical purposes, a detailed lineshape analysis is needed for room temperature studies. In the case of indirect gap PL, this analysis must include the possibility of phonon emission and absorption. The need for a detailed lineshape analysis that yields accurate band gap energies is very acute for the $Ge_{1-y}Sn_y$ system if the purpose of the study is the determination of the crossover composition $y_c$, since small systematic changes in the slope of the compositional dependence of either gap translate into significant changes in the predicted value of $y_c$. In this paper, we present the results of an in-depth study of PL from a large set of $Ge_{1-y}Sn_y$ samples with compositions in the $0 < y < 0.06$ range. A fit of the direct gap emission using a generalized van Roosbroeck-Shockley formula, combined with an accurate model expression for the absorption coefficient, yields values of the direct band gap $E_0$ which in the limit $y \to 0$ are in excellent agreement with the known value for bulk Ge. The compositional dependence of this transition can be determined with meV-level accuracy. For the indirect gap, on the other hand, it is shown that the standard textbook expressions for the absorption coefficient are inadequate for Ge-like materials, especially for $Ge_{1-y}Sn_y$ alloys. As a result of this limitation, the indirect edge energy $E_{ind}$ can only be extracted from the energy of the indirect PL peak by assuming a constant shift that is chosen to make the value of $E_{ind}$ for $y = 0$ agree with the known value of $E_{ind}$ for pure Ge. The crossover composition deduced from these studies is $y_c = 0.073^{+0.007}_{-0.006}$.

**II. EXPERIMENT**

A) SAMPLES

The PL measurements were performed on $Ge_{1-y}Sn_y$ films grown on Si substrates using CVD methods. Some of the samples were deposited using the CVD precursors $Ge_2H_6$ and $SnD_4$, as described in several references.[6,8,19] Other samples, particularly those with high-Sn concentrations, used $Ge_3H_8$ as the Ge-source.[20] Typical thicknesses required for the observation of good PL signals are about 500 nm. A subset of the samples were grown on Ge-buffered Si substrates, and these samples typically display a stronger PL signal, presumably due to the fact that the $Ge_{1-y}Sn_y$/Ge interface is less defected than the $Ge_{1-y}Sn_y$/Si interface. Unfortunately, PL studies of $Ge_{1-y}Sn_y$ films on Ge-buffered samples are difficult for $y \leq 0.02$ due to overlap between the film and buffer signals.

Our study also includes *n*-type samples doped with P using the $P(GeH_3)_3$ precursor.[21-23] The carrier concentrations in the samples were determined from Hall measurements and infrared spectroscopic ellipsometry. The agreement between the two techniques is generally excellent.

The structural properties of the films were monitored using Nomarski microscopy, Atomic Force Microscopy, Rutherford Backscattering (RBS) and x-ray diffraction (XRD). These studies reveal smooth surfaces (RMS roughness ~ 3 nm), very good epitaxial alignment and low defectivity, as evidenced by a drastic reduction of the signal in a channeling geometry. XRD measurements of (224) reciprocal space maps yielded the in-plane and out-of-plane lattice parameters from which the strain could be computed. The 004 ω-rocking curve scans exhibited full width half maxima (FWHM) close to 0.6º, depending on thickness, which were reduced to the 0.1-0.3º range after two or three cycles of Rapid Thermal Annealing at temperatures that were adjusted between 550 ºC and 700 ºC, depending on Sn concentration. The emission from samples with low Sn-concentrations is significantly enhanced after a passivation annealing cycle in an $H_2$ atmosphere.

The Sn concentrations in the films were determined from their relaxed cubic lattice parameter using the compositional dependence of this parameter measured by Beeler *et al.*[24] This dependence was obtained by fitting the measured $a_0$ from tens of $Ge_{1-y}Sn_y$ samples as a function of their Sn concentration determined from RBS. Accordingly, we expect the Sn-concentrations determined from x-ray measurements in our samples to be in close agreement with direct RBS determinations, and this is indeed the case. However, since the x-ray measurements of the lattice parameter are considerably more precise than the RBS fit parameters, plots of the band gap energy versus compositions derived from x-ray measurements are less noisy than equivalent plots using concentrations extracted directly from RBS.

Since the size difference between Ge and Sn is much larger than that between Si and Ge, non-random atomic arrangements are more likely in $Ge_{1-y}Sn_y$ alloys than in their $Ge_{1-x}Si_x$ counterparts. These effects might have an impact on the compositional dependence of optical transitions, and to assess their importance we have carried out detailed experiments to map the Sn distribution in our films. We used electron energy loss spectroscopy (EELS) on a JEOL ARM 200F microscope equipped with a GATAN Enfinium spectrometer. The EELS spectra were collected from 2×2 nm² areas (green box in Figure 1 a) with probe size of 0.12 nm. The sample thickness, determined from low loss spectra, was approximately 60 nm. Multiple scans throughout each sample revealed sharp and well defined ionization peaks corresponding to the characteristic Ge-L and Sn-M core edges at 1217 eV and 483 eV, respectively, indicating that the material is a pure and crystalline $Ge_{1-y}Sn_y$ alloy. The spectral features were then used to create atomic resolution maps for the Sn and Ge constituents shown in Figure 1b-d. In the case of Ge (panel b) we see a uniform pattern of distinct dimer rows in the (110) projection of the cubic lattice in the crystal. The Sn map (panel c) displays a similar arrangement of features corresponding to a two-dimensional projection of the Sn atoms within a sample column with dimensions of 2 nm × 2nm × 60nm probed by EELS. Collectively the maps confirm that the Sn atoms are evenly distributed throughout the Ge matrix and occupy random substitutional sites in the diamond lattice. Finally, the Sn atom map was overlaid onto the Ge map to construct a composite representation of the chemical distribution in the lattice, as shown in panel 1d, which indicates a close alignment of the Ge and Sn dimer rows along the individual (001) columns. We see no diffraction intensity above background levels between the dimer projections rows, indicating that the Sn and Ge atoms occupy the same tetrahedral lattice devoid of precipitates and interstitials.

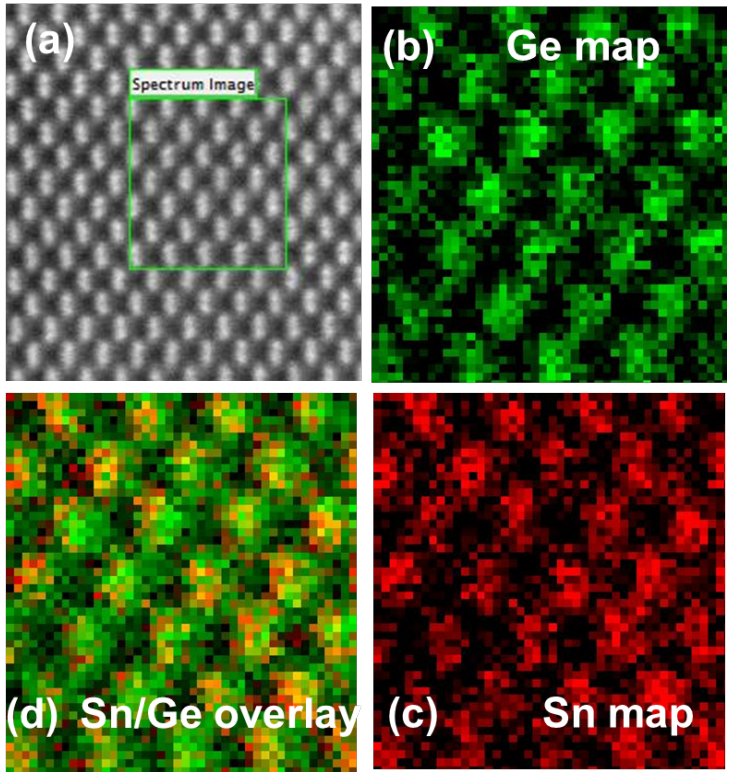

**Figure 1** STEM micrographs and EELS maps of the Ge and Sn atoms in a $Ge_{0.96}Sn_{0.04}$ sample. (a) Survey image of the film identifying the ~ 2 × 2 nm region analyzed by EELS; (b) Ge atom map created from the L edge showing the characteristic dimer rows colored green; (c) Corresponding Sn map colored red was created from the M edge; (d) Overlay of Ge and Sn maps illustrating a uniform distribution of the atomic species over the same average alloy lattice, and a close correspondence of the dimer Ge/Sn rows. The sample thickness used for this analysis was 60 nm.

B) OPTICAL MEASUREMENTS

Photoluminescence was measured at room temperature from samples excited by ~400 mW of cw-980 nm laser radiation, focused to a ~20 μm spot. The emitted light was collected with a Horiba 140 mm f/3.9 Czerny-Turner micro-HR™ spectrometer equipped with a liquid-nitrogen-cooled InGaAs detector. The system response was carefully calibrated using a 10 W tungsten-halogen lamp (Newport Corporation catalog #6318). This calibration is important because of the broad spectral range covered by the observed peaks and the fact that the detector's responsivity drops sharply near 2300 nm. Long-pass filters were used during the measurements to block the PL signal from the Si substrate and the laser radiation, which appears as a strong peak at 1960 nm due to second-order diffraction from the 600 gr/mm grating. In many spectra, a residual laser peak is seen at 1960 nm and subtracted from the data by fitting it with a Gaussian profile. From the width of the Gaussian we determine the spectral resolution of the measurements (FWHM), which is found to be 16 meV.

## III. RESULTS

Figure 2 shows photoluminescence spectra for a few selected samples. The spectra consist of a main peak, assigned to the direct band gap, and a weaker peak at lower energy that is assigned to the indirect gap. Both peaks shift to lower energy as the Sn-concentration is increased, but their separation decreases. The solid lines in Fig. 2 correspond to a fit that is discussed below. At concentrations above 5.5% the two emission peaks appear completely merged.

## IV. THEORY

The spontaneous emission transition rate $R$ per unit sample volume, for photons with energy $E$ emitted into solid angle $d\Omega$, is given by[25,26]

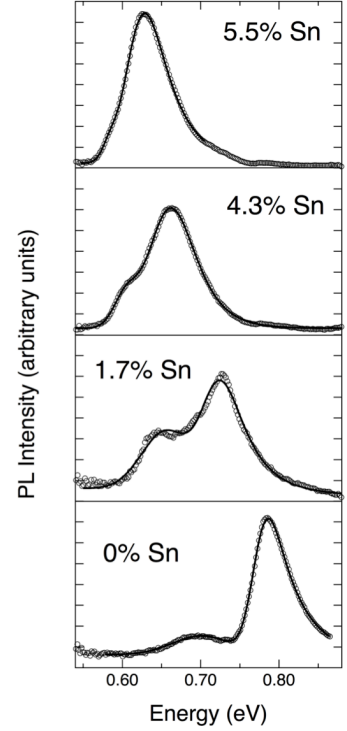

**Figure 2** Room temperature photoluminescence spectra from representative $Ge_{1-y}Sn_y$ films. The main peak is assigned to direct gap emission, and the weaker feature at lower energies corresponds to indirect gap emission. The solid lines show fits using a simple Gaussian for the indirect gap and an Exponentially Modified Gaussian for the direct transition.

$$\frac{1}{V}\frac{d^2 R}{d\Omega dE} = \frac{2}{h^3}\left(\frac{n_{op} E}{c}\right)^2 \left[\frac{\alpha(E)}{\exp\left(\frac{E-\Delta F}{k_B T}\right)-1}\right] \quad (1)$$

In this generalized van Roosbroeck-Shockley expression we use the standard notation for fundamental physical constants, $T$ for the temperature, and $n_{op}$ for the material's index of refraction. Here it is assumed that in a sample under steady-state photoexcitation, quasi-equilibrium conditions are reached separately in the conduction and valence bands, and we define

$$\Delta F = E_{Fc} - E_{Fv}, \quad (2)$$

where $E_{Fc}$ ($E_{Fv}$) is the quasi-Fermi level in the conduction (valence) band. A self-absorption correction can be included by multiplying the theoretical prediction times a factor $\{1-\exp[-\alpha(E)L]\}/\alpha(E)L$, where $L$ is the sample thickness. This assumes that the photoexcited carriers are uniformly distributed over the volume of the film, a reasonable expectation in view of the very high ambipolar diffusion coefficient in Ge.[27,28] A more realistic expression that includes surface and interface recombination was deemed unnecessary for our purposes because the self-absorption- related shifts in the fitted gap energies are never higher than 1 meV. Only the high-energy tail of the emission is moderately affected by self-absorption, as expected.

The absorption coefficient $\alpha(E)$ that appears in Eq. (1) results from the difference between the absorption and stimulated emission rates. In the case of vertical (direct) transitions between isotropic parabolic bands, it is given by[26]

$$\alpha(E) = \alpha_0(E)\left[f_v(E) - f_c(E)\right], \quad (3)$$

where $\alpha_0(E)$ is the absorption coefficient for empty conduction bands and full valence bands, and we use the Fermi functions

$$f_c(E) = \frac{1}{\exp\{[E_c(E) - E_{Fc}]/k_B T\} + 1}$$
$$f_v(E) = \frac{1}{\exp\{[E_v(E) - E_{Fv}]/k_B T\} + 1} \quad (4)$$

Here $E_c(E)$ and the $E_v(E)$ are the energies of the electrons in the conduction and valence band states, respectively, vertically separated by an energy $E$.

For the empty-band absorption coefficient $\alpha_0(E)$ we developed an analytical model, including excitonic effects, that is discussed in full detail in Refs. 6, 7, and 18. Our analytical expression reproduces the bulk Ge absorption curve over a range of up to 0.1 eV above the direct band gap based on standard band structure parameters, without introducing any additional parameter to adjust for the absorption strength. It also lends itself to its use in $Ge_{1-y}Sn_y$ alloys by extrapolating the relevant band structure parameters from bulk Ge using $\mathbf{k \cdot p}$ expressions. A comparison of the emission lineshape predicted by this theory with simpler textbook models is presented in the Appendix.

**Table I** Deformation potentials used to calculate strain effects on the direct and indirect gap transitions in $Ge_{1-y}Sn_y$ alloys.

| Direct gap hydrostatic deformation potential $a_{E0}$ (in eV)[a] | Indirect gap hydrostatic deformation potential $a_{ind}$ (in eV)[b] | Valence band shear deformation potential $b$ (in eV)[c] |
|---|---|---|
| -9.46+2.33y | -3.6+1.77y | -1.88 eV |

[a]References 26 and 28, as discussed in text.
[b]References 27, 29, and 30, as discussed in text.
[c]Reference 31.

The effect of strain is easily incorporated in our modeled absorption using deformation potential theory. Table I shows the deformation potentials used for these calculations. For pure Ge, the hydrostatic deformation potential for the direct gap is obtained by plotting the pressure dependence of $E_0$, measured by Goñi *et al*, (Ref. 29) as a function of the relative volume change $\Delta V/V_0$, and fitting a linear function up to $\Delta V/V_0 = 2.5\%$. For the indirect gap, we use experimental data from Ahmad and Adams.[30] For α-Sn, there are no experimental measurements of deformation potentials. For the $E_0$ gap, Li *et al.*(Ref. 31) calculated the hydrostatic deformation potential for both Ge and α-Sn. For Ge, the theoretical value must be multiplied times a factor 0.94 to match the experimental value. We then multiply the theoretical value for α-Sn by the same factor to obtain the deformation potential we use for this material. For the indirect gap, we follow the same procedure, but this time we use theoretical data from Schmid *et al.*(Ref. 32) and Brudevoll *et al.*(Ref. 33). For the alloy, we interpolate linearly between Ge and α-Sn. For the shear deformation potential we are not aware of measurements or calculations for α-Sn, so we use the result $b = -1.88$ eV for Ge obtained by Liu *et al.*[34] from photoreflectance measurements in tensile-strained Ge layers. More recently, Lin *et al.*[35] presented photoreflectance measurements from compressively strained $Ge_{1-y}Sn_y$ alloys from which they obtain a value $b = -4.07 \pm 0.91$ eV, very different from the Liu *et al.*[34] result. Indirect evidence suggests that the Liu *et al* value is more accurate. For example, in addition to $b$ one can also extract the *hydrostatic* deformation potential from the photoreflectance measurements in strained films. In the case of Liu *et al*,[34] the agreement with the pressure data from Goñi and co-workers[29] is nearly perfect, whereas the hydrostatic deformation potential from Lin *et al.*[35] is about 15% larger than the value in Table I. Theoretical calculations, for example Ref. 36, are also closer to the Liu *et al*. value. A significant difference between the Liu and Lin experiments is that the former were obtained in films under tensile strain, whereas the latter correspond to compressive strain. This suggests that a careful comparative study of the elastic properties of Ge under tensile or compressive strain is warranted. In our case, since most of our samples experience a slight tensile strain after annealing, we use the values from Liu *et al.*[34]

A fit of the direct gap emission using Eq. (1) combined with Eq. (3) and (4) requires three adjustable parameters: the direct gap energy $E_0$, the absorption broadening $w_{abs}$, and the photoexcited carrier density $n_{ex}$. The strain is obtained from the x-ray experiments and the effective masses are determined from ***k·p*** expressions as a function of the $E_0$ parameter. The photoexcited charge density $n_{ex}$ is used to determine the quasi-Fermi levels that appear in Eq. (2) using standard textbook expressions. For the conduction band, our simulation includes the two valleys around the Γ and L points, and for the valence band it includes the light- and heavy hole bands. The fits reported here assume parabolic bands, but we have verified that the inclusion of non-parabolicity effects does not change the value of the $E_0$ parameter. In the case of doped n-type samples, we add the doping concentration to the parameter $n_{ex}$. Adjusting $n_{ex}$ shifts the peak emission as described in the appendix, and affects the high-energy tail in the emission spectra. For the levels of absorption broadening needed to fit our data, the emission lineshape is quite insensitive to the value of $n_{ex}$ if $n_{ex} < 10^{18}$ cm$^{-3}$. The highest value that we obtained from our fits was $n_{ex} = 6 \times 10^{19}$ cm$^{-3}$, but most samples are fitted with $n_{ex}$ values in the $10^{17}$ - $10^{18}$ cm$^{-3}$ range. The temperature $T$ that appears in the theoretical expressions was taken as $T = 316$ K. This value was obtained by performing a few preliminary fits in samples measured at increasing laser power densities, for which we observed a downshift in the $E_0$ energy. If we attribute this shift to laser heating, we obtain from the known temperature dependence of the direct gap in Ge that for the laser power used for our measurements the temperature is raised by about 23 K. In some cases, particularly in the thinnest samples, the predicted high-energy tail is somewhat broader than the experimental one even at the lowest photoexcited carrier concentrations. We do not think this is due to inaccuracies in the self-absorption correction, which, if significant, are more likely to manifest themselves in thick samples. Instead, since the high-energy tail is exponentially sensitive to the temperature, we believe that sample heating in thin films may be somewhat less than assumed, because part of the incident laser light is absorbed in the buffer layer.

Eq. (1) is also valid for indirect transitions,[37] but Eq.(3) is not, and therefore the calculation of the absorption coefficient is less straightforward. Moreover, textbook expressions for the indirect absorption coefficient $\alpha_0(E)$ are of the form[38-41]

$$\alpha_0 = K \left[ \frac{(E + E_{phon} - E_{ind})^2}{\exp(E_{phon}/k_B T) - 1} + \frac{(E - E_{phon} - E_{ind})^2}{1 - \exp(-E_{phon}/k_B T)} \right], \quad (5)$$

where $E_{phon}$ is the energy of the phonon involved and the factor $K$ consolidates several constants and slowly-varying functions of $E$. Among these is a factor $(E_0 - E)^{-2}$ that originates from the denominator of the second-order perturbation theory expression that is required to compute the indirect absorption as a phonon-assisted process. Here $E_0$ represents the average energy of the dominant intermediate state, very close to the direct gap $E_0$, as indicated in the inset of Fig. 3. In Ge-like materials, however, the $(E_0 – E)^{-2}$ factor is not slowly varying but strongly resonant, as seen in Fig. 3, where we plot Eq. (5) with $K$ = constant and with $K$ = constant/ $(E_0-E)^2$. The reason for this behavior is that the direct gap $E_0$ and the indirect gap $E_{ind}$ are very close in energy. In $Ge_{1-y}Sn_y$ alloys the denominator is even more resonant, and Eq. (5) should be expected to break down completely. The energy denominator cannot be taken out of the density of state integrations, and therefore one no longer obtains the well-known quadratic numerators in Eq. (5). It is also important to point out that even if a more rigorous expression is developed to replace Eq. (5), the inclusion of excitonic effects at room temperature is likely to be important in a quantitative theory of indirect gap emission, in much the same way that excitonic effects are crucial to explain the shape and strength of the direct gap absorption at room temperature.[6,7] Moreover, to the extent that $Ge_{1-y}Sn_y$ alloys deviate more strongly from a virtual crystal approximation than $Ge_{1-x}Si_x$ alloys, one cannot rule out a strong no-phonon contribution to the indirect gap emission, which would render the theory significantly more complex. Due to these complications, the development of a quantitative rigorous theory of indirect gap spontaneous emission in $Ge_{1-y}Sn_y$ alloys is beyond the scope of this paper, but the lack of appropriate theoretical expressions creates a challenge when it comes to extracting values of the indirect gap from the measured PL spectra. While the direct gap component of the spectra in Fig. 2 can be fitted using Eq. (1) combined with Eq. (3) and the analytical expression for the absorption coefficient in Refs. 6 and 7, the lack of an equivalent expression for the indirect gap emission makes it impossible to fit the entire PL spectrum consistently using physically motivated lineshape expressions.

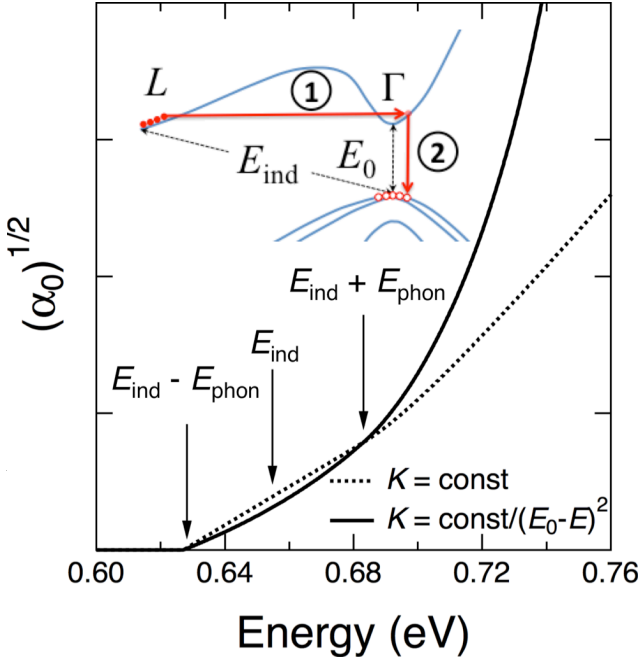

**Figure 3** Calculated indirect gap absorption for pure Ge using Eq. (5) with $K$ = constant (dotted line) and with $K$ = constant/$(E_0-E)^2$ (solid line). For easy comparison, the relative scale of the plots was adjusted so that the curves coincide at $E = E_{ind}+E_{phon}$. The large discrepancy between the two expressions, which increases in $Ge_{1-y}Sn_y$ alloys, indicates a breakdown of the standard theory that leads to Eq. (5). The inset shows the relevant portion of the electronic band structure and the two-step emission process leading to the resonant denominator. Here (1) is a phonon-assisted transition from the $L$-valley to the $\Gamma$ valley, followed in (2) by recombination with a hole in the valence band.

Due to the theoretical limitations, the indirect gap emission component is simply modeled as a Gaussian. Of course, the peak of the Gaussian cannot be identified with the indirect gap, but we assume that it is shifted from $E_{ind}$ by a constant amount, independent of composition, that can be determined from PL experiments in pure Ge films for which the indirect band gap is known. By using this procedure, and assuming that for the very modest values of strain in our samples the indirect emission is dominated by transitions to the heavy-hole band, and that the indirect band gap of pure Ge at 316 K is 0.655 eV,[42] we find that the constant shift that must be applied to the data is 0.031 eV. This value is consistent with the energies of the zone edge LA/LO (0.027/0.031 eV) and TO (0.035 eV) phonons in Ge. The expected compositional dependence of these phonon energies is rather weak for $y < 0.1$, which justifies the use of the same rigid shift of 0.031 eV for all $Ge_{1-y}Sn_y$ samples.

Direct fittings of the data using Eq. (1) plus the Gaussian lineshape for the indirect gap are very difficult due to the poor convergence of our microscopic expression for the direct gap. To address this problem, we developed a two-step-approach, described in detail in the Appendix, in which we start by fitting *both* edges with empirical expressions, namely a Gaussian for the indirect gap, and a so-called so-called Exponentially Modified Gaussian (EMG)[43-45] for the direct gap. The EMG is the convolution of a Gaussian and an exponential decay, and its adjustable parameters are the center and width of the Gaussian and the decay constant of the exponential. In the second step we fit the EMG component with the theoretical expression for the direct gap. One important advantage of this approach is that the absorption broadening can be easily extracted from the Gaussian width parameter of the EMG, so that the number of adjustable parameters in the second step of the fit, in which Eq. (1) is used, is reduced to just $E_0$ and $n_{ex}$. Moreover, as shown in the Appendix, the (fortuitous) agreement between the EMG lineshape and the actual theoretical emission is so good that there is no loss in the ultimate accuracy of the $E_o$ gap extracted from the two-step procedure.

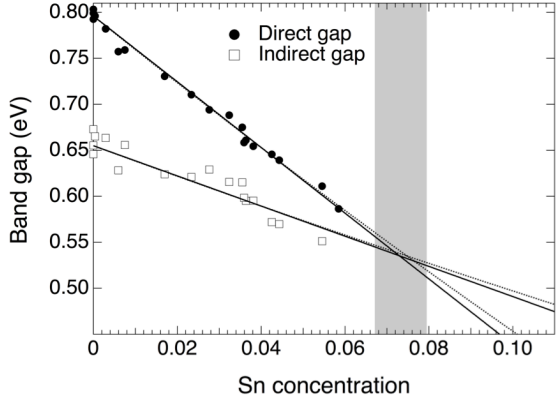

**Figure 4** Compositional dependence of the direct gap $E_0$ (circles) and the indirect gap $E_{ind}$ (squares) in $Ge_{1-y}Sn_y$ alloys. The solid lines represent linear fits to the data, and the dotted lines correspond to quadratic fits for the entire compositional range $0 \leq y \leq 1$ using the end values at $y = 0$ and $y = 1$ as fixed parameters. The shaded area around the crossover point indicates the error in the crossover concentration.

An initial test of our fitting procedure is the extraction of the direct band gap of pure Ge from PL experiments on Ge-on-Si films. At the assumed temperature $T = 316$ K, the accepted value of the direct band gap of Ge is $E_0 = 0.796$ eV,[18] whereas our fits for the pure Ge films gives $E_0 = 0.798 \pm 0.005$ eV. We thus believe that our method makes it possible to determine the band gap $E_0$ within a few meV. It is important to point out here that since the calculated direct gap absorption deviates strongly from the expected $(E-E_0)^{1/2}$ dependence for free electron-hole pairs (due to excitonic effects), the difference between the band gap $E_0$ and the maximum of the calculated emission cannot be expected to be $k_BT/2$, as commonly assumed. This is shown in detail in the Appendix.

## V. DISCUSSION

### A) COMPOSITIONAL DEPENDENCIES

Fig. 4 shows results for all $Ge_{1-y}Sn_y$ samples for which the indirect band gap emission was observed, and we see that the data is significantly noisier for the indirect gap. We believe, however, that this higher noise is not due to the *ad hoc* nature of the indirect emission lineshape, but to the fact that the indirect band gap emission is considerably weaker, and sometimes nearly merged, with the stronger direct gap emission.

The simplest fits to the data in Fig. 4 are linear expressions given by (in eV):

$$E_0(y) = 0.796 - (3.57 \pm 0.06)y$$
$$E_{ind}(y) = 0.655 - (1.64 \pm 0.10)y \quad (6)$$

The fits are shown as solid lines in Fig. 4. Here we fix the independent term to match the known Ge values. If we make both coefficients adjustable, we obtain similar results (in eV) [ $E_0(y) = (0.792 \pm 0.002) - (3.47 \pm 0.09)y$ and $E_{ind}(y) = (0.662 \pm 0.005) - (1.82 \pm 0.16)y$ ]. As expected from the introduction, extrapolation of Eq. (6) to $y = 1$ leads to values drastically different from those measured in α-Sn, so that the linear expression in Eq. (6) can only be valid over the range of our measurements. To extend the results to the entire compositional range, it is usually assumed that the compositional dependence is quadratic. The quadratic term is proportional to a bowing parameter $b$ defined from

$$E_0(y) = E_0^{Ge}(1-y) + E_0^{Sn}y - b_0 y(1-y)$$
$$E_{ind}(y) = E_{ind}^{Ge}(1-y) + E_{ind}^{Sn}y - b_{ind} y(1-y) \quad (7)$$

If both end values for $y = 0$ and $y = 1$ are taken as equal to the corresponding values in Ge and α-Sn, respectively, we are left with a fitting function that contains the bowing coefficient as its only adjustable parameter. Using $E_0^{Ge} = 0.796$ eV, $E_0^{Sn} = -0.413$ eV (Ref. 46), $E_{ind}^{Ge} = 0.655$ eV, and $E_{ind}^{Sn} = -0.035$ eV (Ref. 47), we obtain $b_0 = 2.46 \pm 0.06$ eV and $b_{ind} = 0.99 \pm 0.11$ eV. One potential difficulty with this approach is that α-Sn is not stable at room temperature. Its low temperature properties were measured in the 1970's, but extrapolation to room temperature is problematic. α-Sn films that were metastable beyond room temperature were grown in the 1980's on InSb and CdTe substrates.[48,49] Extensive work was done on the optical properties of the metastable films,[50] but the low-energy band structure has not been revisited. The value $E_0^{Sn} = -0.413$ eV that we used for our $E_0$ fit was determined by Groves *et al.* from magnetoreflection experiments at low temperatures.[46] This value is in good agreement with theoretical calculations based on semi-empirical methods.[33,51] No temperature dependence was found below 100K, but assuming the same value at room temperature may introduce an unknown error. The indirect gap was determined at low temperatures using free-carrier reflectivity[52] ($E_{ind} = 0.115$ eV) and electrical measurements[47] ($E_{ind} = 0.092$ eV). The agreement with theory is not particularly good. Chelikovsky and Cohen predict $E_{ind} = 0.140$ eV (Ref. 51), whereas Brudevoll *et al* find $E_{ind} = 0.175$ eV.(Ref. 33) The indirect transition has a relatively strong temperature dependence that has been measured only to 250 K in Ref 52. Extrapolating the results to room temperature, we obtain $E_{ind}$ = -0.035 eV, and this value was used to obtain the bowing parameter. Had we used the theoretical value $E_{ind}^{Sn}$ = 0.175 eV, the bowing parameter fit would have been $b_{ind}$ = 1.2±0.1 eV. These results illustrate the need for specifying the end values when comparing bowing parameters from different authors. For this reason, the simple fit in Eq. (6) may be preferable.

Earlier PL work by our group on samples with $y \leq 0.03$ indicates $b_0$ = 1.8 eV, significantly below the value reported here.[11] The band gap values in Ref. 11 were assumed to be equal to the PL maximum. Given the far more sophisticated lineshape analysis in the present paper, the extended compositional range, and the much larger data set, we believe that the results reported here are more accurate than those in Ref. 11. On the other hand, $E_0$ measurements using spectroscopic ellipsometry and photoreflectance[5] also yield $b_0$ = 1.94±0.04 eV. However, this value corresponds to an extended compositional range $y \leq 0.14$. The ellipsometric data are in very good agreement with the PL data in Fig. 4 over the overlapping range $y \leq 0.06$. The different bowing parameters obtained over different compositional ranges suggest a possible compositional dependence of the bowing parameter itself, *i.e.* a deviation from the quadratic dependence in Eq. (7). This is not entirely surprising, since the functional form of Eq. (7) can only be justified theoretically for a very weak alloy perturbation to the virtual crystal, and it is by no means obvious that $Ge_{1-y}Sn_y$ is close to this limit. However, Pérez Ladrón de Guevara and co-workers obtain $b_0$ = 2.3±0.1eV, in better agreement with our PL results, from absorption measurements on seven samples with $y \leq 0.14$ grown by rf sputtering on Ge substrates.[53] The apparent discrepancy with Ref. 5 can be traced back to the measured band gaps at the highest Sn concentrations. For example, for $y$ = 0.14, Pérez Ladrón de Guevara *et al.* find $E_0$ = 0.33 eV, whereas for the same concentration D'Costa *et al* find $E_0$ = 0.41 eV. It is difficult to compare the results from the two references, since the samples were grown using a different method, the compositions were determined differently, and the band gaps were extracted from the optical data using a different methodology. Most importantly, however, the determination of possible deviations from Eq. (7) will require a very large set of samples covering a compositional range beyond that in Fig. 4.

In Ref. 35, Lin *et al* presented photoreflectance results from four samples with $y < 0.064$. The reported bowing parameter is $b_0$ = 2.42±0.04 eV, in excellent agreement with our PL data. Chen *et al.* also reported photoluminescence results from five samples in the compositional range $0 < y < 0.09$. The band gap is simply identified with the PL peak maximum. The resulting bowing parameter is $b_0 = 2.1\pm0.1$ eV. Given the different methodology and reduced sample set, the agreement with our data is satisfactory.

Experimental work on the compositional dependence of the indirect band gap is much more scarce. Tonkikh *et al* report low-temperature PL results from samples in the $0.06 \leq y \leq 0.09$ range.[54] They obtain a bowing parameter $b_{ind} = 0.80\pm0.06$ eV by assuming that the photoluminescence peak corresponds to the no-phonon line, and $b_{ind} = 0.17\pm0.06$ eV if they assume that the peak corresponds to emission of an LA phonon. Recent PL experiments in GeSn quantum wells by the same group suggest instead a lower limit $b_{ind} > 1.47$ eV.[55] Our fit value is intermediate between the different bowing parameters proposed by Tonkikh *et al*. Mathews *et al*. (Ref. 11) did not give an explicit compositional dependence of the indirect transition, but their measured dependence of the $E_0$-$E_{ind}$ difference suggests a direct-indirect cross-over near $y_c = 0.09$, somewhat higher than the cross-over value discussed below.

The failure of the early theoretical predictions regarding the compositional dependence of the direct gap has motivated a renewed theoretical interest in $Ge_{1-y}Sn_y$ alloys. Using 64-atom quasi-random structures to simulate the alloy, Yin *et al*. predict $b_0 = 2.55$ eV and $b_{ind} = 0.89$ eV, in good agreement with our results.[56] Their calculated direct-indirect crossover concentration is $y_c = 0.063$. Using smaller (16 atom) cells, Chibane and Ferhat[57] predict a composition-dependent bowing parameter close to $b_0 = 2.9$ eV for $y = 0.06$ and approaching $b_0 = 1.9$ eV for $y = 0.19$. For the indirect gap, Chibane and Ferhat predict $b_{ind} = 0.9$ eV, leading to $y_c = 0.105$. The strong compositional dependence of $b_0$ is in contrast with the results from Yin *et al*., who find a weak compositional dependence for this parameter.[56] A non-monotonic compositional dependence of $E_0$ is predicted by Lee *et al* based on even smaller 8-atom cells.[58] Using similar cells but with an empirical pseudopotential method, Moontragoon *et al*. (Ref. 59) predict $b_0 = 2.49$ eV and $b_{ind} = 2.28$ eV, which lead to $y_c = 0.17$. Perhaps the most important conclusion from a comparison of all these theoretical predictions with our experimental data is that the best agreement is obtained for 64-atom supercells, suggesting that this is probably the minimum supercell size for meaningful comparisons with experiment. It is interesting to point out in this context that an empirical correction to the VCA was introduced by Gupta and coworkers.[60] Here a local disorder correction is added to the VCA pseudopotential. This correction includes a free parameter that is adjusted to reproduce the experimental bowing coefficient for the direct gap, which is taken as $b_0$ = 2.1 eV. The calculations predict $b_{ind}$ = 0.91 eV, which is close to our experimental value. Unfortunately, the predictions from this approach have not been compared with the experimental bowing coefficients for higher-energy transitions,[5] which might provide a more complete assessment of the validity of this intriguing method.

B) CROSSOVER CONCENTRATION

Since the difference in slopes for the two lines in Fig. 4 is not too large, small slope errors translate into a relatively large uncertainty regarding the crossover concentration $y_c$ between the direct and indirect band gaps. From the slopes determined above and their uncertainties, we obtain $y_c = 0.073^{+0.007}_{-0.006}$. In principle, the crossover concentration can be determined directly from low-temperature PL or from measurements of the infrared Drude reflectivity in doped samples. However, these experiments are challenging because the electronic density of states is much larger at the $L$ valley minimum than at the $\Gamma$ minimum. In the case of doped samples, for example, simulations show that the predicted Drude response only reflects the lower effective mass of carriers at the $\Gamma$ minimum at concentrations much higher than $y_c$ = 0.073, for which there is a substantial population of the $\Gamma$ valley. It is also possible that the alloy perturbation will partially mix the conduction band states when they overlap in energy, further blurring the abruptness of the transition.

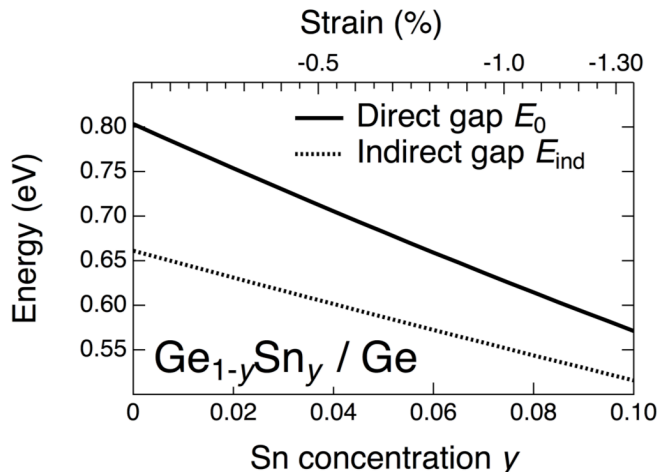

**Figure 5** Predicted compositional dependence of the direct and indirect edges in $Ge_{1-y}Sn_y$ alloys fully strained to Ge substrates. The two transitions were calculated using the results in Eq. (10) combined with standard deformation potential theory.

The results obtained can be used to estimate the compositional dependence of the direct and indirect edges in $Ge_{1-y}Sn_y$ films fully strained to Ge substrates. Figure 5 shows the predicted dependence using the fits in Fig. 4 and standard deformation potential theory, with parameters from Table I. Compressive strain increases the separation between the direct and indirect edges, which is predicted to be 0.076 eV at the crossover composition $y_c = 0.073$ for the relaxed alloy. If the calculation is continued to higher Sn-concentrations, a direct-indirect crossover is predicted for fully-strained $Ge_{1-y}Sn_y$ at $y_c = 0.19$, close to the value $y_c = 0.17$ estimated by Tonkikh et al.[54] However, the compressive strain value at such composition would be 2.7%. At this level of strain both the linear elasticity and deformation potential theories underlying the calculation will break down, so that the estimate may be affected by a significant systematic error. However, regardless of the precise crossover value, it is apparent that fully strained $Ge_{1-y}Sn_y$ direct gap semiconductors would be of very limited practical interest.

C) DOPED SAMPLES

The effect of phosphorus doping on the band gap energies of Ge has been somewhat controversial as of late. A clear band gap renormalization was reported by Haas in 1962 (Ref. 61) for both $E_0$ and $E_{ind}$, and reproduced theoretically.[62-65] More recently, however, room temperature PL experiments on highly doped Ge films on Si were interpreted as showing a negligible $E_0$ renormalization.[14] Subsequent PL work, on the other hand, showed clear evidence for an $E_0$ renormalization shift that is comparable to Haas' earlier work.[18,66,67] Our results make it possible to extend the study of band gap renormalization effects to $Ge_{1-y}Sn_y$ alloys.

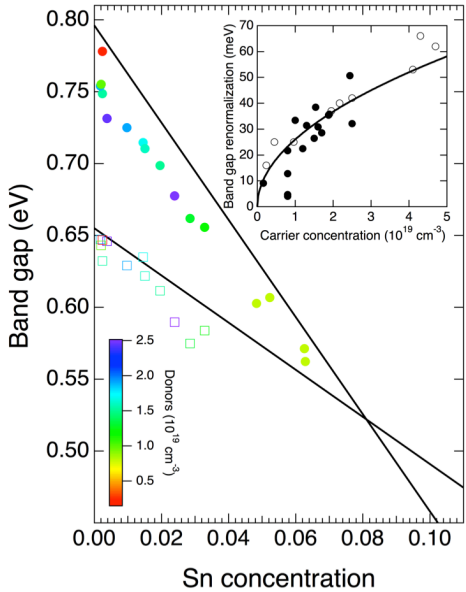

**Figure 6** Direct gap $E_0$ (circles) and indirect gap $E_{ind}$ (squares) obtained from the PL spectra of n-type $Ge_{1-y}Sn_y$ films. Colors indicate the carrier concentration. The solid lines are the best fits for the direct and indirect gap in undoped films, from Fig. 4.
The inset shows the band gap renormalization extracted from the data (black circles). The white circles are results for bulk Ge from Ref. 61. The solid line is a fit to the data as described in the text.

In Fig. 6 we compare the measured values of $E_0$ and $E_{ind}$ in our doped $Ge_{1-y}Sn_y$ films with the corresponding results for undoped samples, shown as solid lines extracted from the best-fit lines in Fig. 4. It is apparent that the energies of both direct and indirect edges are reduced in the presence of donor atoms, as expected from theory. The band gap renormalization can be extracted by simply subtracting the measured values in doped samples from the best prediction for the band gap of an undoped sample with the same Sn concentration. In principle, this renormalization will depend on the Sn concentration, but since the electronic structure of our Ge$_{1-y}$Sn$_y$ alloys is not too different from that of pure Ge, we expect the doping concentration dependence of the renormalization to be much stronger than its compositional dependence. This seems to be corroborated by the plot in the inset in Fig. 6, where we show the renormalization energy as a function of the doping concentration, regardless of the samples' Sn concentrations, and we observe a clear trend. The empty circles in the inset show data for pure Ge, and we see that the two data sets are consistent. The Ge$_{1-y}$Sn$_y$ data are somewhat noisier but this is fully expected because of fluctuations in the undoped gap value on the order of those seen in Fig. 4. The solid line in the inset shows a fit of the data with the function $\Delta E_0 = -A\sqrt{n/10^{18}\text{cm}^{-3}}$, with $A$ = 8.22 meV. We obtain a somewhat better agreement with the data using a linear fit, as proposed in Ref. 67, but the square-root function gives the correct dependence for $n \to 0$. It is important to point out that our analysis of the PL results includes the possible effects of band filling and associated shifts of the peak emission, so that our results represent a measurement of the true band gap renormalization. Finally, the agreement of the renormalization shifts measured in our Ge$_{1-y}$Sn$_y$ alloys with those reported in literature for pure Ge provide additional indirect evidence for the accuracy of the underlying compositional dependence of $E_0$, which is used to compute the renormalization shifts. Moreover, we note that the samples whose $E_0$ energies have the largest negative deviations from the solid line in Fig. 4 correspond to the highest values of $n_{ex}$ in our fits, suggesting these deviations may not be random fluctuations but reflect a band gap renormalization contribution, so that the actual noise in the data may be even smaller than suggested by Fig. 4. On the other hand, for the case of the indirect gap the extracted renormalization shifts are too noisy to identify a clear trend beyond a reduction in band gap energy with doping.

## VI. CONCLUSION

In summary, we have presented a study of the compositional dependence of the direct and indirect band gap edges in Ge$_{1-y}$Sn$_y$ alloys using room temperature photoluminescence. This technique is in principle the most attractive one for the determination of the two gaps, because they produce distinct features in the same spectrum. However, a detailed theoretical analysis is needed to extract the needed gap values from the broadened room-temperature spectra, and this need is particularly acute in the $Ge_{1-y}Sn_y$ system because small errors in the compositional dependence of either gap leads to a large uncertainty in the value of direct-indirect crossover. We have developed the required accurate model for the direct gap transition and shown that it allows for the determination of the energy $E_0$ with meV precision. On the other hand, we have shown that current models of indirect gap emission are inadequate for Ge-like $Ge_{1-y}Sn_y$ alloys. The indirect gap energies were extracted from the peak energy of the emission by subtracting a constant energy that was found to be approximately equal to the energy of the phonons involved in this transition. Further theoretical work will be needed to improve this aspect of the analysis and fit the indirect emission with a realistic physical model of the process.

The results presented here for both direct and indirect transitions indicate a crossover concentration $y_c = 0.073^{+0.007}_{-0.006}$. This is substantially less than predicted from theoretical models within the virtual crystal approximation. Explicitly incorporating alloy effects via large supercells brings the theoretical predictions much closer to the experimental data. The low value of the predicted crossover concentration indicates that direct-gap $Ge_{1-y}Sn_y$ can be easily fabricated using current growth approaches. However, the very different density of states between the $L$ and $\Gamma$-valleys in the conduction band of $Ge_{1-y}Sn_y$ semiconductors suggest that optical and electrical behavior of the alloys may not reflect their direct-gap character until the Sn concentration is significantly higher than $y_c$.

## VII. ACKNOWLEDGMENTS

This work was supported by the Air Force Office of Scientific Research under contracts DOD AFOSR FA9550-12-1-0208 and DOD AFOSR FA9550-13-1-0022.

**APPENDIX**

In this appendix we provide details on our theoretical calculation of the emission rate and on the EMG functions used in the intermediate data processing steps. The solid line in Fig A1 shows the emission obtained from Eq. (1) and (3) using our model for $\alpha_0$,[6, 7] calculated for the case of pure Ge with a photoexcited carrier concentration $n = 5 \times 10^{18}$ cm$^{-3}$. As indicated in Ref. 7, the best agreement between calculated and experimental absorption in Ge and $Ge_{1-y}Sn_y$ alloys is obtained using Gaussian broadening. Accordingly, we broaden the absorption $\alpha_0$ by convoluting the calculated function with a Gaussian with a FWHM $w_{abs}$. A simpler expression of the form $K(E-E_0)^{1/2} \exp[-(E-E_0)/k_B T]$, where $K$ is a constant, appears in most textbooks [25,26,40,41] and is often applied to the analysis of PL data. The simplified expression follows from Eq. (1) under the assumption of Boltzmann statistics and direct recombination of free electron-hole pairs from parabolic energy bands. It is easy to show that the maximum of this function is located at $E_{max} = E_0 + k_B T/2$, independent of the difference $\Delta F$ of quasi-Fermi levels, which only appears in the energy-independent factor $K$. On the other hand, we show in Fig. A2 the difference $E_{max}-E_0$ as a function of $\Delta F$ for the theoretical lineshape in Fig. A1. For low values of $\Delta F$, $E_{max}-E_0$ is independent of $\Delta F$, as in the simplified expression above, but the separation $E_{max}-E_0$ is closer to $k_B T/4$. This is directly related to the inclusion of excitonic effects in the absorption model. As shown in Ref. 6, the rise of the absorption coefficient above the direct gap $E_0$ is much steeper when the effect of excitons is accounted for, and this shifts the peak of Eq. (1) towards the band gap $E_0$. The error incurred by using the standard $k_B T/2$ correction is about 6 meV at room temperature, which is small compared with the accuracy with which band gaps are known in semiconductors, but amounts to a not entirely negligible 4% of the direct-indirect separation in pure Ge, and an even larger fraction of this separation in $Ge_{1-y}Sn_y$. Using $E_0 = E_{max}-k_B T/4$, on the other hand, gives excellent direct gap values directly from the experimental data without the need of any fit, but becomes less accurate in the presence of strain due to the different weights of the contributions from the split heavy- and light-hole bands.

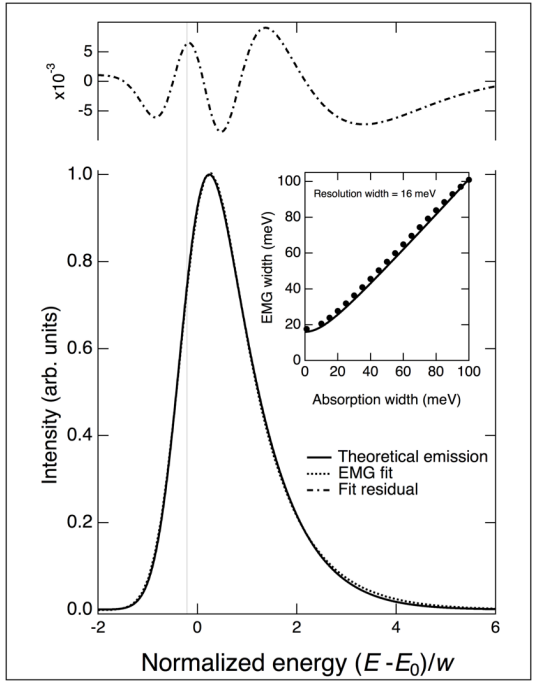

**Figure A1**  EMG fit of a theoretical direct gap emission lineshape. The horizontal axis is given in terms of the normalized energy $(E-E_0)/w$, where $w$ is given in Eq. (8). The vertical grey line indicates the value of the "location" parameter of the EMG, which corresponds to the center of its Gaussian component. The inset shows the Gaussian width fit parameter from the EMG (circles) from fits of theoretical lineshapes with different absorption broadenings. The solid line is from Eq. (8).

For $\Delta F > 0.45$ eV, the peak energy in Fig. A2 shifts to higher values, reflecting band filling effects. The value $\Delta F = 0.45$ eV corresponds to a carrier concentration $n \sim 10^{17}$ cm$^{-3}$. Larger band filling effects are computed for Ge$_{1-y}$Sn$_y$ alloys due to the reduced separation between the direct and indirect minima in the conduction band. Fig. A2 could be used to read off the direct band gap values from the observed $E_{max}$ values in unstrained Ge, except that one needs to take into account the finite spectrometer resolution, which leads to an additional upshift because the theoretical profile is asymmetric. For the conditions of our experiment, where the resolution is about one-half of the absorption broadening, the upshift is ~1.3 meV, which is rather small.

Since our model expressions for direct gap emission account realistically for excitonic, strain, and band-filling effects, the accuracy of the $E_0$ gaps determined with the method used here is considerably improved. However, as indicated above, we adopted an intermediate step in which we fit the experimental data with an EMG profile, and therefore we need to investigate the quality of these fits to determine if our fitting process could induce systematic energy shifts. The dotted line in Fig. A1 is a fit of the theoretical emission using an EMG, and we see that the agreement is extremely good. (In the practical application we use the reverse procedure, in which we fit the raw data with an EMG, followed by a fit of the EMG with Eq. (1), but the quality of the fit is of course identical).

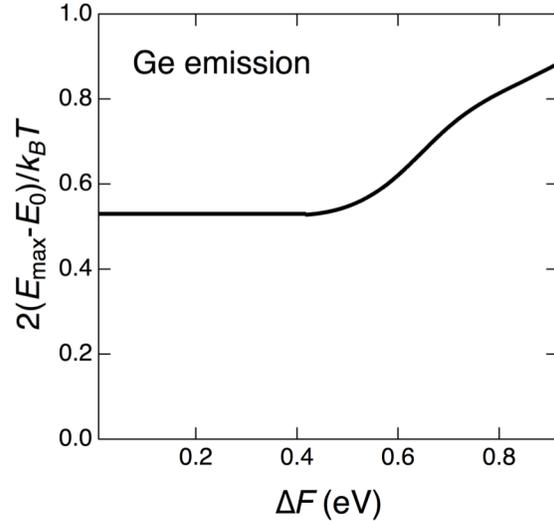

**Figure A2**  Separation between the location of the maximum $E_{max}$ of the theoretical direct gap emission and the band gap value $E_0$, in units of $k_B T/2$ for $T = 300$K as a function of the difference in quasi-Fermi levels between the conduction and valence bands.

We repeated the fits for theoretical emission lineshapes computed for different values of the absorption broadening (FWHM) $w_{abs}$ and a fixed instrumental broadening (FWHM) $w_{res}$ = 0.016 eV, as in our experiments. All fits are of similar quality as the one depicted in Fig. A1. The inset shows the width parameter $w$ (FWHM) of the Gaussian component of the EMG, plotted as a function of $w_{abs}$. The solid line corresponds to the expression

$$w = \sqrt{w_{abs}^2 + w_{res}^2} \;, \tag{A1}$$

and we see that it is in excellent agreement with the measured values (although in principle it should only be exactly valid for the convolution of two Gaussians). Thus the parameter $w_{abs}$, needed for the computation of the emission profile using Eq. (1) can be extracted from Eq. A1 using the widths $w$ from the EMG fits. This eliminates one of the adjustable parameters of the fit with Eq (1), as discussed above.

The quality of the approximation in Eq. A1 decreases in the presence of strain, because our absorption calculation includes the separate light- and heavy hole edges explicitly, but we still use a single EMG function in the first step. In these cases we find that the fit of the EMG with the theoretical expression for the emission can be further improved by slightly reducing $w_{abs}$ from the value obtained from Eq. A1.

In view of the remarkable agreement between EMG profiles and those predicted from Eq. (1), it could be argued that a "first principles" theoretical fit is not needed, and that the desired band gap could be extracted from the location parameter of the EMG. Unfortunately, this is not the case. The horizontal axis in Fig. A1 is normalized so that the zero corresponds to the energy of the direct gap $E_0$. The vertical grey line shows the "location" parameter in the EMG fit (corresponding to the location of the Gaussian component). We see that that the two values do not agree, and both are shifted from $E_{max}$. These relative shifts depend in a complicated way on the other fit parameters, making it virtually impossible to extract reliable $E_0$ values from the EMG parameters. Only when the EMG is fitted with a physically motivated expression is it is possible to determine $E_0$ with the required accuracy. Similar considerations apply to the indirect gap emission. It could be argued that assuming a simple Gaussian in the first step of our fit represents an oversimplification. In fact, we have explored a considerably more sophisticated approach, in which we model the indirect emission with three EMG's (for phonon emission, phonon absorption, and no-phonon transitions), and we obtain excellent fits that appear to indicate that the no-phonon transitions increase in relative intensity as a function of the Sn concentration. However, we have not been able to find a reliable way to extract the desired $E_{ind}$ value from these fits in the absence of a theoretical model of indirect emission as available for the direct gap. The "location" parameter of the EMG functions is not related to the desired $E_{ind}$ in any obvious way and has a complex interaction with the width parameters. On the other hand, the simple Gaussian curve has the virtue that position and width are completely "decoupled", and therefore the location of this Gaussian is a far more robust route to the desired $E_{ind}$.

The treatment of broadening in the calculated emission is important for the accuracy of the band gap fit parameter $E_0$. As indicated above, we broaden the absorption $\alpha_0$ by convoluting the calculated function with a Gaussian with a FWHM $w_{abs}$. We calculate the emission in Eq. (1) using the broadened absorption, and then we convolve this calculated emission with a Gaussian with FWHM $w_{res}$ to account for the instrumental resolution. In earlier work[18] we used a simplified approach in which we calculated the absorption without any broadening and performed a single Gaussian convolution after calculating the spontaneous emission. This step was supposed to account effectively for the absorption broadening and instrumental resolution. However, we have noticed that this leads to band gap values that are systematically shifted to lower energies by about 10 meV. The two-step broadening procedure described here, on the other hand, is more physically appealing and leads to excellent agreement with the expected band gap in the case of pure Ge films.